\documentclass[twocolumn]{autart}    

\usepackage{graphicx}          
 \usepackage{mathtools}
 \usepackage{amssymb}
 \usepackage{balance}
 \usepackage{float}
\usepackage[scr]{rsfso}
 
 \newtheorem{assumption}{Assumption}
  \newtheorem{definition}{Definition}

\begin{document}

\begin{frontmatter}

\title{Consistency Analysis of the Simplified Refined Instrumental 
Variable Method for Continuous-time Systems\thanksref{footnoteinfo}} 

\thanks[footnoteinfo]{This paper was not presented at any IFAC meeting.}

\author{Siqi Pan}\ead{siqi.pan@uon.edu.au},      
\author{Rodrigo A. Gonz\' alez}\ead{grodrigo@kth.se},	
\author{James S. Welsh}\ead{james.welsh@newcastle.edu.au} \;and  
\author{Cristian R. Rojas}\ead{crro@kth.se}  



\begin{abstract}                          
In this paper, we analyse the consistency of the Simplified Refined Instrumental 
Variable method for Continuous-time systems (SRIVC). It is well known that the 
intersample behaviour of the input signal influences the quality and accuracy of the 
results when estimating and simulating continuous-time models. Here, we present a 
comprehensive analysis on the consistency of the SRIVC estimator while taking into 
account the intersample behaviour of the input signal. The main result of the paper 
shows that, under some mild conditions, the SRIVC estimator is generically consistent. 
We also describe some conditions when consistency is not achieved, which is important 
from a practical standpoint. The theoretical results are supported by simulation examples.
\end{abstract}

\end{frontmatter}

\section{Introduction}		\label{sec:intro}

Direct continuous-time (CT) identification algorithms based on sampled data 
have achieved remarkable success in many practical 
applications~\cite{Young2011,Young2006a,Garnier2008,Garnier2014}. 
In particular, the Refined Instrumental Variable method for Continuous-time systems (RIVC) and 
its simplest embodiment, the Simplified RIVC (SRIVC) \cite{Young1980} are considered to be the 
most reliable algorithms in CT system identification~\cite{Garnier2008,Garnier2014} and have been 
used in practice for almost 40 years. However, there has been limited theoretical 
support for these algorithms. Most of the discussions with respect to the properties of these 
estimators, such as consistency and statistical efficiency, are based on empirical 
observations~\cite{Young1980,Young2011,Garnier2008,Young2008,Young2002,Young2006}. 
By providing the user with the theoretical properties of the estimator, they will be better informed 
as to the conditions necessary to achieve accurate estimates. It is therefore important for any estimator 
used in practice to have solid theoretical support. 

The objective of this paper is to provide the theoretical support in terms of consistency for the 
SRIVC estimator. Based on the theoretical results, the user will then be able to make conscious decisions when 
applying this algorithm in practice with respect to obtaining consistent estimates. 
There have been some attempts in the existing literature~\cite{Liu2011,Chen2017,Young2015} to examine the 
consistency property of the SRIVC estimator. However, they are all based on the well-developed 
theoretical results in discrete-time (DT) system 
identification~\cite{Soderstrom1983,Soderstrom1981,Stoica1983,Stoica1981}, which do not provide 
a mechanism to include the intersample behaviour of the underlying CT system.
The current paper analyses the consistency property of the SRIVC estimator while taking into account the 
intersample behaviour of the signals.
The theoretical results we obtain here explicitly inform the user when consistency can be achieved 
by the SRIVC estimator. The conditions under which the estimator 
is not consistent are clearly stated as well, and suggestions are also given to alleviate the bias on the estimates 
in practical situations. 
We would like to reiterate that if it is important, in practice, to have a consistent estimator, then the theoretical 
results presented in this paper clearly describe how this can be achieved.

As mentioned above, a significant challenge presented when estimating a CT system is that only sampled 
input-output data are available as measurements. Therefore, the measured input needs to be interpolated in 
some manner in order to simulate a CT model output.
It has been discussed in \cite{Schoukens1994,Andersson1994} that violating the input intersample 
behaviour assumption of the underlying data generating process may lead to severe estimation errors.
A simple motivation example has been provided in \cite{Schoukens1994} to illustrate the modelling 
error induced in the estimation process when a band-limited input is assumed while the true system input 
is a zero-order hold.
In some practical situations, the input intersample behaviour will be unknown, e.g. environmental modelling, 
and some interpolation methods have been discussed in~\cite{Soderstrom2009} to better approximate the 
input in these cases.
We note that if the input is applied and controlled by the user, then it generally can be interpolated exactly.
There are, however, some exceptions, e.g. identification of cascaded systems. 
The input signal to the second system in a cascaded system will be
a continuous function of time that cannot be interpolated exactly between samples. Hence, an 
error will be induced on the modelled output, and this in turn will affect the estimated parameters.
Thus, it is important to take into account the intersample behaviour of the signals 
when dealing with CT estimators, which has been somewhat overlooked 
in the existing literature.
We also note that there are some CT identification methods that do not require the 
sampled signals to be interpolated. For example, higher order Pad\'e approximation is used in~\cite{Chou1999}
to approximate a DT filter that produces the same output as the sampled CT filter output, 
which avoids the need to reconstruct the CT input signal, and CT models are identified 
in~\cite{Marelli2010} based on second-order statistics.


The consistency analysis in the current paper has some 
similarities to the work in~\cite{Liu2011} as both analyses are based on the 
consistency theorem found in~\cite[Theorem~4.5]{Soderstrom1983} developed for the DT bootstrap 
instrumental variable (IV) method. The work in~\cite{Liu2011} analyses the convergence of the 
RIVC estimator with an autoregressive noise model. 
There are, however, a 
few shortcomings associated with the analysis in~\cite{Liu2011}. An extra filter is introduced  for the 
purpose of discretising the derivatives of the input signal, which is not part of the RIVC implementation. We note 
that this extra filter introduces unnecessary complexity into the analysis, and its role can be replaced by filters that 
are already part of the algorithm. In addition, due to the formulation of the proof in~\cite{Liu2011}, 
non-causal filters may arise since the system and model are allowed 
to be parameterised as biproper transfer functions. 
Furthermore, the first step to establish the convergence, and therefore the consistency of the RIVC algorithm is to 
show that a solution of the estimator exists~\cite{Liu2011}. This relies on 1)~the Sylvester matrices constructed 
from the system 
and model polynomials being non-singular, and 2)~the expectation of the two filtered input vectors, 
denoted by~$\Phi$ in~\cite{Liu2011}, being non-singular. 
Firstly, in~\cite{Liu2011}, the non-singularity of the Sylvester matrix does 
not comply with their given assumptions. The analysis assumes that the system and model are parameterised with 
monic denominator polynomials. It can be shown that this implicit assumption, together with Assumption~A5 
in~\cite{Liu2011} on the model order, results in the Sylvester matrix constructed from the system polynomials 
being singular when the degree of the model denominator is greater than that of the system. The proof of 
Theorem~1 in~\cite{Liu2011} with respect to assumption~A5 therefore cannot proceed once the Sylvester matrix 
is singular. 
Secondly, it is stated in~\cite{Liu2011} that showing the non-singularity of $\Phi$ relies on the matrix
$E\{UU^\top\}$ being non-singular, where $U$ is the vector containing the input samples with sample size $N$ 
(see (19) in~\cite{Liu2011}). 
We note that $E\{UU^\top\}$ has dimension $N\times N$, and is only non-singular up to the order of persistent 
excitation of the input. Thus, in the asymptotic case, it is not sufficient to conclude that~$\Phi$ is always non-singular 
under the persistent excitation assumption given in~\cite{Liu2011}.


Other work related to consistency such as~\cite{Chen2017} assumes that the model structure is exactly known 
and does not take into account the intersample behaviour of the input as part of the analysis. 
The work in~\cite{Young2015} describes a unified Refined Instrumental Variable (RIV) approach for estimating 
DT or CT transfer functions characterised by a unified operator that can be interpreted in 
terms of a backward shift, derivative or delta operator. This unified RIV~\cite{Young2015} is 
suggested to be optimal in maximum likelihood, prediction error minimisation and instrumental variable 
terms under the Box-Jenkins model structure for both discrete and continuous-time. 
However, only limited theoretical analysis is provided with respect to the consistency of the estimates 
by using an incremental implementation of the algorithm with no explicit mention of the 
intersample behaviour of the signal. 
By neglecting the intersample behaviour as part of the analysis, the results in~\cite{Liu2011,Young2015,Chen2017} 
have overlooked the possibility that the converging point of the estimator no longer corresponds to the true 
system parameters when the system input cannot be interpolated exactly.

In this paper, we analyse the consistency property by incorporating the intersample behaviour of the 
input, the output and the instrument signals. The main result of the paper shows that the SRIVC estimator 
is generically consistent under some mild conditions 
in the presence of additive coloured noise on the measured output.
In the proof of the consistency theorem, the use of an additional 
filter, such as in~\cite{Liu2011}, is avoided by discretising the derivatives of the input signal with filters that are 
already part of the SRIVC implementation. The denominator of the model is parameterised as a non-monic 
polynomial. This ensures that the Sylvester matrix is always non-singular for model orders satisfying the condition 
for the existence of a unique solution, i.e. when the model orders are equal to the system orders, or when one of 
the model polynomial degrees is greater than that of the system.
Two common interpolation methods, namely the first-order hold (FOH) and the zero-order hold (ZOH), are 
considered for the true system input to conduct the consistency analysis in the main theorem and the corollaries.
In addition, in the first part of the consistency theorem where the existence of a solution is shown, we employ 
the notion of generic consistency~\cite[Theorem~4.1]{Soderstrom1983}, i.e. the set of normal matrices 
of the SRIVC method yield inconsistent estimates has Lebesgue measure zero. 
This implies that there are rare cases where a 
certain combination of the input and system parameters can make the normal matrix singular even though all the 
assumptions are satisfied. We show the generic non-singularity of the normal matrix through the use of analytic 
functions by following the method presented in~\cite[Lemma~1]{finigan1974}.
Furthermore, we have shown that the intersample behaviour of the input in the 
instrument vector does not influence the consistency of the SRIVC estimator; however, in order for the SRIVC 
estimator to be generically consistent, the intersample behaviour of the input in the regressor vector must match 
that of the true system input. We also note that the intersample behaviour of the output 
in the regressor vector does not impact on the consistency of the SRIVC estimator at the converging point 
of the iterative algorithm. 
The practical implications of this paper are to inform the user of the conditions necessary to 
achieve consistency of the SRIVC estimator and also of the conditions that may lead to an inconsistent estimate.

This paper is organised as follows. Section \ref{sec:pre} provides system and model definitions as well as 
a description of the SRIVC estimator and the definition of generic consistency. 
This is followed by Section \ref{sec:main}, where the theoretical results 
of the paper, including the consistency theorem and its related corollaries and remarks, are presented. 
Section \ref{sec:sim} provides simulation results that support the theoretical analysis, and the paper is 
concluded in Section \ref{sec:conclusion}.


\section{Preliminaries}	\label{sec:pre}

In this section, we define the structure of the continuous-time single-input single-output system and 
model and provide a brief description of the SRIVC estimator and the definition of generic consistency.

\begin{figure} [h]
\begin{center}
\includegraphics[width = 6.5cm]{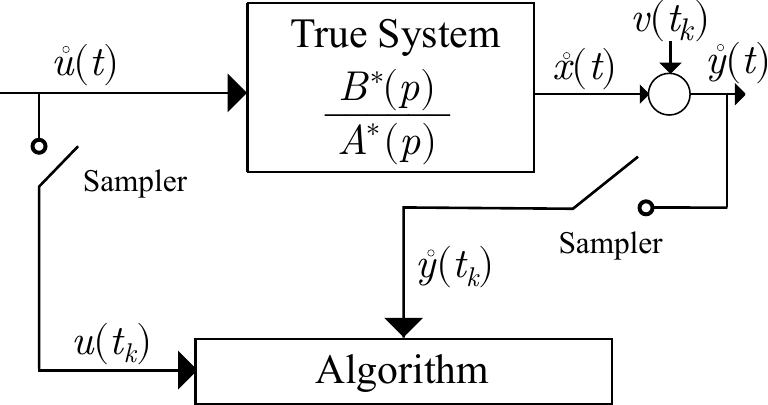}
\caption{Continuous-time system including the sampling and signal notations.}
\label{fig:true_sys}
\end{center}
\end{figure}

The true system is described as a proper transfer function given by
\begin{equation} \label{eq:sys}
  S: \begin{cases}
    \mathring{x}(t) &= \frac{B^*(p)}{A^*(p)}\mathring{u}(t)	 \\
    \mathring{y}(t) &= \mathring{x}(t) + v(t),
  \end{cases}
\end{equation}
where the circles ($\mathring{.}$) signify that the signals are associated with the true system. 
The numerator and denominator polynomials are coprime with degrees given by $m^*$ and 
$n^*$ respectively, i.e.
\begin{equation}	\label{eq:true_par}
\begin{split}
	B^*(p) &= b_0^*p^{m^*} + b_1^*p^{m^*-1} + \cdots + b_{m^*}^*	\\ 
	A^*(p) &= a_1^*p^{n^*} + a_2^*p^{n^*-1} + \dots + a_{n^*}^*p + 1
\end{split}
\end{equation}
with $p$ being the differential operator, i.e. $p^{i}x(t) = \frac{d^{i} x(t)}{dt^i}$. The additive 
noise on the output is coloured and expressed as
\begin{equation}
	v(t) = H(p)e(t),
\end{equation}
where $H(p)$ is an inversely stable filter and $e(t)$ a zero-mean Gaussian noise, i.e. $e \sim N(0,\sigma^2)$.
The output observation equation of the CT system~\eqref{eq:sys} at sample instance~$t_k$ 
is given by
\begin{equation*}
	\mathring{y}(t_k) = \mathring{x}(t_k) + v(t_k),
\end{equation*}
where $\mathring{x}(t_k)$ is the unobserved, noise-free output. 
It is well known that CT white noise does not have a finite variance \cite{Astrom1970}, which makes 
computing its 
time-derivatives particularly difficult. Due to this difficulty and the DT nature of the sampled signals, 
we only consider DT noise in this paper. The true system and the nature of sampling are shown in 
Fig.~\ref{fig:true_sys}.

The model is also parameterised as a proper transfer function
\begin{equation} \label{eq:model}
  M: \begin{cases}
    x(t_k) &= \frac{B(p)}{A(p)}u(t_k)	\\
    y(t_k) &= x(t_k) + e(t_k),
  \end{cases}
\end{equation}
where $u(t_k) = \mathring{u}(t)$ at the sampling instants, and the numerator polynomial with
degree $m$ and the denominator polynomial with degree $n$ are given by
\begin{equation}	\label{eq:model_parameters}
\begin{split}
	B(p) &= b_0p^{m} + b_1p^{m-1} + \cdots + b_m	\\
	A(p) &= a_1p^{n} + a_2p^{n-1} + \dots + a_np + 1.
\end{split}
\end{equation}
The unknown parameter vector is then defined as
\begin{equation*}
	\theta = \left[\begin{array}{cccccc}
	a_1 & \dots & a_n & b_0 & \dots & b_m
	\end{array}\right]^\top.
\end{equation*}
Note that in the sequel when a mixed notation of CT operators and DT data is 
encountered in the analysis such as in~\eqref{eq:model}, it implies that the input~$u(t_k)$ 
in~\eqref{eq:model} is interpolated in some manner, e.g. using either a zero-order or a first-order hold, 
and the resultant output is sampled at $t_k$.

The SRIVC estimator minimises the sum of squares of the generalised equation error (GEE) \cite{Young1981} 
$\varepsilon(t_k)$, which is given by
\begin{align} \label{eq:gee}
	\varepsilon(t_k) &= \mathring{y}(t_k) - x(t_k) 	\notag \\
				& = \mathring{y}(t_k) - \frac{B(p)}{A(p)}u(t_k)	\notag \\
				&= \frac{1}{A(p)}(A(p)\mathring{y}(t_k) - B(p)u(t_k))	\notag \\
				&= A(p)y_f(t_k) - B(p)u_f(t_k),
\end{align}
where
\begin{equation} \label{eq:prefilter}
	y_f(t_k) = \frac{1}{A(p)} \mathring{y}(t_k), \; \text{and} \; u_f(t_k) = \frac{1}{A(p)} u(t_k).
\end{equation}
Due to the iterative nature of the SRIVC method, the $(j+1)$-th iteration of the SRIVC estimate \cite{Young1980,Garnier2008} based on parameters estimated in the $j$-th iteration is given by
\begin{equation}	\label{eq:srivc}
	\theta_{j+1} = \left[\frac{1}{N} \sum_{k=1}^N \hat{\varphi}_f(t_k) \varphi_f^\top(t_k) \right]^{-1} 
	 \left[\frac{1}{N} \sum_{k=1}^N \hat{\varphi}_f(t_k) y_f(t_k) \right],
\end{equation}
where
\begin{align}		\label{eq:srivc_reg}
	\varphi_f(t_k)
	&= \frac{1}{A_j(p)} \Big[\begin{array}{ccc}
	-p^n\mathring{y}(t_k) & \dots & -p\mathring{y}(t_k) 
	\end{array} \notag \\
	&\;\;\;\;\;\;\;\;\;\;\;\;\;\;\;\;\;\;\;\;\;\;\;\;\;\;\;\;\;\;\;
	\begin{array}{ccc}
	p^mu(t_k) & \dots & u(t_k)
	\end{array}\Big]^\top, 	
\end{align}
and
\begin{align}		\label{eq:srivc_ins} 
	\hat{\varphi}_f(t_k) &= \left[
	-x_f^{(n)}(t_k)  \dots  -x_f^{(1)}(t_k) \; u_f^{(m)}(t_k)  \dots  u_f(t_k) \right]^\top\notag \\
	&= \frac{1}{A_j(p)} \Big[\begin{array}{ccc}
	-\frac{B_j(p)}{A_j(p)} p^nu(t_k) & \dots & -\frac{B_j(p)}{A_j(p)}pu(t_k)
	\end{array}	\notag \\
	&\;\;\;\;\;\;\;\;\;\;\;\;\;\;\;\;\;\;\;\;\;\;\;\;\;\;\;\;\;\;\;
	\begin{array}{ccc}
	p^mu(t_k) & \dots & u(t_k)
	\end{array}\Big]^\top. 	
\end{align}
The algorithm is stopped either when a maximum number of iterations is reached or when the 
relative error between the previous and current estimate is smaller than a prefixed constant, i.e.
\begin{equation}	\label{eq:srivc_stop}
	\frac{\|\theta_{j+1}-\theta_j\|}{\|\theta_{j+1}\|} < \epsilon.
\end{equation}

Next, we provide a definition of generic non-singularity~\cite{Soderstrom1983} and relate it to the 
definition of generic consistency.

\begin{definition}
Consider an $n\times n$ matrix $R(\rho)$, which depends on a finite-dimensional vector $\rho$. Then, 
$R$ is generically non-singular with respect to $\rho$ if the set $\{\rho: rank R(\rho)<n\}$ has Lebesgue 
measure zero.
\end{definition}

\begin{definition}
The SRIVC estimator \eqref{eq:srivc} is generically consistent if the term in the matrix inverse
in \eqref{eq:srivc} is generically non-singular, and for all $j\geq 1$, the set of parameter
values for which the estimates do not converge to the true parameters as the sample size tends 
to infinity has Lebesgue measure zero.
\end{definition}

\begin{figure} [h]
\begin{center}
\includegraphics[width = 8.5cm]{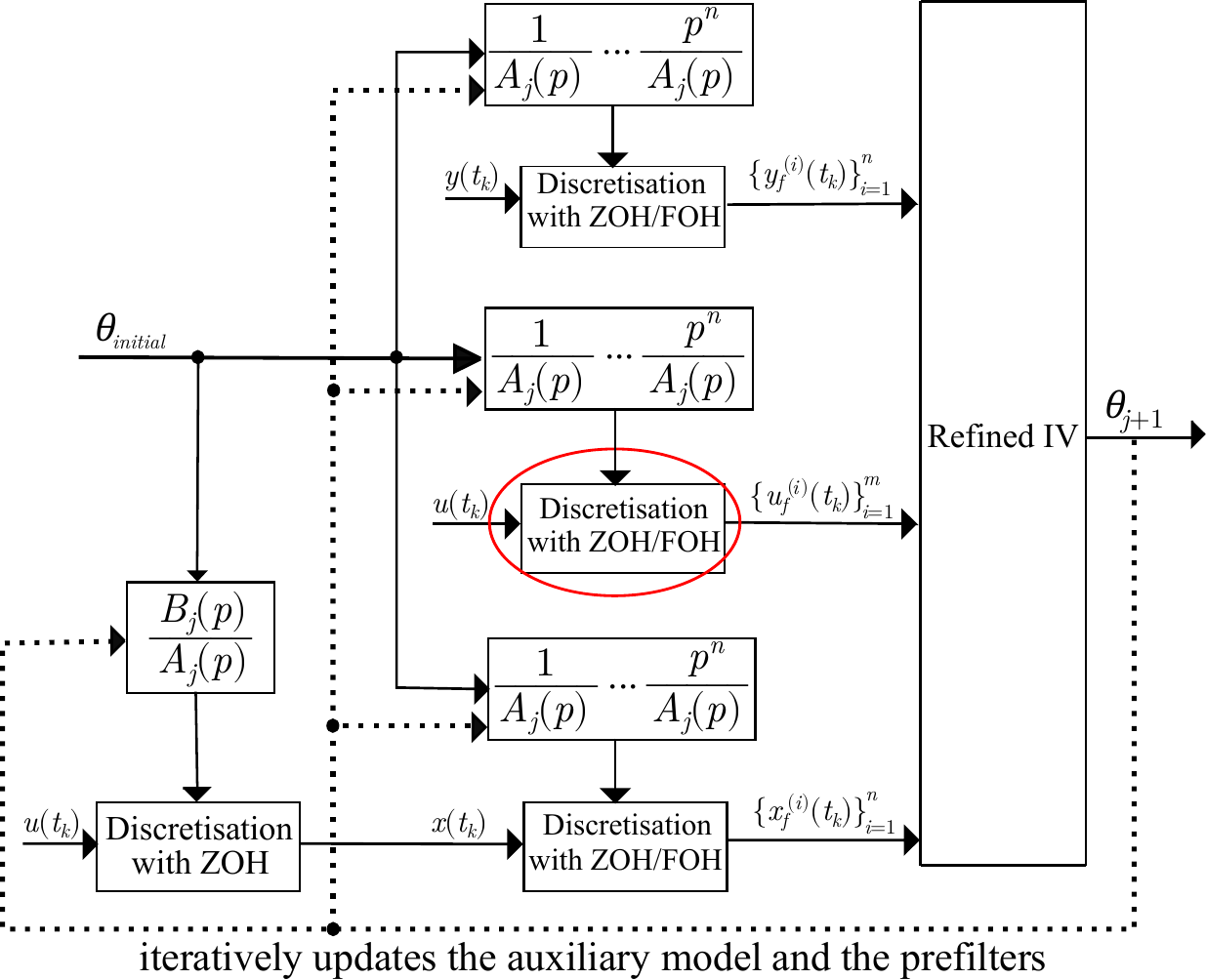}
\caption{Implementation of the SRIVC algorithm. Note that we use the notation $u^{(n)}(t_k)=p^n u(t_k)$.}
\label{fig:SRIVC}
\end{center}
\end{figure}

We note that all the filtering operations are performed in discrete-time within the implementation of 
the SRIVC estimator, hence the need to explicitly consider the intersample behaviour of the 
signals in any analysis. A block diagram depicting the SRIVC algorithm is shown in 
Fig.~\ref{fig:SRIVC}. In the following section, we investigate the effect of the intersample 
behaviour of the sampled data on the consistency of the SRIVC estimator. It will be shown that 
the input signal in the regressor vector, i.e. the model input, is required to have the same intersample 
behaviour as the input applied to the true system for the SRIVC estimator to be generically consistent. This 
intersample behaviour is circled in Fig.~\ref{fig:SRIVC}.

\section{Theoretical Results}		\label{sec:main}

In this section, we develop a theorem that establishes the consistency of the SRIVC estimator, as well 
as some corollaries and remarks that examine the consistency with respect to different intersample conditions.
Some additional lemmas required by the proof of the consistency theorem are presented in the Appendix.

For simplicity, the analysis will be presented for the case of a single-input single-output (SISO), linear, time 
invariant (LTI), asymptotically stable system with regularly sampled data. We note that the 
analysis can be easily extended to multi-input single-output (MISO) systems, though the extension may 
be difficult for multi-input multi-output (MIMO) systems. 
\vspace{-0.3cm}
We next state the assumptions required in Theorem~\ref{consistency} for the SRIVC estimator to be 
generically consistent.
\begin{assumption}	\label{assum1}
The true system $\frac{B^*(p)}{A^*(p)}$ is proper 
($n^* \geq m^*$) and asymptotically stable with $A^*(p)$ and $B^*(p)$ being coprime.
\end{assumption}
\vspace{-0.3cm}
\begin{assumption}	\label{assum2}
The input $u(t_k)$ and disturbance $v(t_s)$ are stationary and mutually independent for all 
$k$ and $s$.
\end{assumption}
\vspace{-0.3cm}
\begin{assumption}	\label{assum3}
The input $u(t_k)$ is persistently exciting of order no less than $2n+1$.
\end{assumption}
\vspace{-0.3cm}
\begin{assumption}	\label{assum4}
All the zeros of $A_j(p)$ have strictly negative real parts, $n \geq m$, with $A_j(p)$ and $B_j(p)$ 
being coprime.
\end{assumption}
\vspace{-0.3cm}
\begin{assumption}	\label{assum5}
The degrees of the polynomials in the model satisfy 
$\min(n-n^*,m-m^*) = 0$.
\end{assumption}
\vspace{-0.3cm}
\begin{assumption}	\label{assum6}
The intersample behaviour of the input $\mathring{u}(t)$ applied to the true system is known 
exactly.
\end{assumption}
\vspace{-0.3cm}
\begin{assumption}	\label{assum7}
The sampling frequency is more than twice of the largest imaginary part of the zeros of $A_j(p)A^*(p)$.
\end{assumption}

When estimating the parameters of a transfer function, unstable zeros in the denominator polynomial 
may arise within the iterations of the algorithm. A simple way to deal with this is to reflect the unstable zeros 
by the imaginary axis. Hence, Assumption~\ref{assum4} is commonly satisfied in practice.
Note that Assumption~\ref{assum5} ensures a unique solution for the model 
parameters~\cite{Soderstrom1983}.
Also note that in the proof of Theorem~\ref{consistency}, we assume the intersample behaviour in 
Assumption~\ref{assum6} to be a FOH to conduct the analysis. We could have equally 
chosen to use a ZOH to model the intersample behaviour of the input as shown in 
Corollary~\ref{cor_zoh}. Furthermore, the prefilters for the output are assumed to be discretised 
with the same hold as the intersample behaviour of the input, i.e. a FOH in Theorem~\ref{consistency}. 
Assumption~\ref{assum7} avoids the problem of aliasing and ensures a meaningful model to be obtained 
according to the Shannon-Nyquist theorem.
Next, we present the main theorem of the paper on the generic consistency of the SRIVC estimator.

\begin{thm} [Generic consistency] \label{consistency}

Consider the SRIVC estimator described in \eqref{eq:srivc}, and suppose Assumptions 
\ref{assum1}-\ref{assum7} hold. Then, for a first-order hold (FOH) input, the following statements are true:
\begin{enumerate}
\item The matrix $E\{\hat{\varphi}_f(t_k) \varphi_f^\top(t_k)\}$ is generically non-singular.
\item The true parameter $\theta^*$ is the unique converging point.
\item As the sample size $N$ approaches infinity, $\theta_{j+1}$ in~\eqref{eq:srivc} 
converges to $\theta^*$ for $j \geq 1$.
\end{enumerate}
\end{thm}


\begin{pf*} {\textit{Proof of Theorem~\ref{consistency}, Statement~1.}}
By substituting
\begin{equation*}	 \label{eq:out}
	\mathring{y}(t_k) = \frac{B^*(p)}{A^*(p)} u(t_k) + v(t_k)
\end{equation*}
into \eqref{eq:srivc_reg}, we can express the regressor vector as
\begin{equation} \label{eq:phif2}
	\varphi_f(t_k) = \frac{1}{A_j(p)A^*(p)}Q_{reg}
				 - \frac{1}{A_j(p)} P_{reg},
\end{equation}

\vspace{-0.7cm}
where

\vspace{-0.7cm}
\begin{equation*}
\begin{split}
	Q_{reg} &= \Big[\begin{array}{ccc}
	-p^nB^*(p)u(t_k) & \dots & -pB^*(p)u(t_k)
	\end{array}	\\
	&\;\;\;\;\;\;\;\;\;\;\;\;\;\;\;\;\;\;\;\;\;\;\;\;\;\;\;
	\begin{array}{ccc}
	p^mA^*(p)u(t_k) & \dots & A^*(p)u(t_k)
	\end{array}\Big]^\top,
\end{split}
\end{equation*}

\vspace{-0.7cm}
and

\vspace{-0.7cm}
\begin{equation}	\label{eq:P_reg}
	P_{reg} = \left[\begin{array}{cccccc}
		p^n v(t_k) & \dots & p v(t_k) & 0 & \dots & 0
		\end{array} \right]^\top.
\end{equation}
The highest order derivative in $Q_{reg}$ that satisfies Assumption~\ref{assum5} is 
$\text{max}(n+m^*,n^*+m) = n+m$.
The vector $Q_{reg}$ can then be expressed as a product of an $(n+m+1)\times(n+m+1)$ Sylvester matrix 
and a vector containing the derivatives of $u(t_k)$, i.e.
\begin{equation*}
	Q_{reg} = S(-B^*(p),A^*(p))U_{du}
\end{equation*}
where
\begin{equation}	\label{eq:syl1}
S(-B^*(p),A^*(p)) = \left[\begin{array}{ccccccc}
-b_0^*	&	-b_1^*	&	\cdots 	& -b_{m^*}^*	&	0		&			&	0	\\
		&	\ddots   	&			&				&	\ddots	& \ddots 	&	       \\
0		&			&	-b_0^*	&	-b_1^*		&	\cdots	&	-b_{m^*}^*	&	0	\\	\hline
a_1^*	&	a_2^*	&	\cdots	&	a_{n^*}^*	&	1		&			&	0	\\
		&	\ddots   	&			&				&	\ddots	& \ddots 	&	      \\
0		&			&	a_1^*	&	a_2^*		&	\cdots	&	a_{n^*}^*	&	1	\\	
\end{array}\right],
\end{equation}
and
\begin{equation}		\label{eq:Udu}
	U_{du} = \left[ \begin{array}{cccc}
	u^{(n+m)}(t_k)  & u^{(n+m-1)}(t_k)  & \dots & u(t_k)
	\end{array} \right]^\top.
\end{equation}
The Sylvester matrix given in~\eqref{eq:syl1} 
is non-singular when $B^*(p)$ and $A^*(p)$ are coprime~\cite{Soderstrom1983}. We 
require~\eqref{eq:syl1} to remain non-singular under the three conditions imposed by 
Assumption~\ref{assum5}, i.e. 1)~the order of the true system is known exactly, 2)~the numerator of 
the model is overfitted, and 3)~the denominator of the model is overfitted. 

We note that there are $n$ rows of the numerator coefficients and $m+1$ rows of the denominator 
coefficients. Now, condition~1 corresponds 
to~\eqref{eq:syl1}. Under condition~2, when the numerator polynomial is overfitted, i.e. 
$m-m^*=l$ for all positive integers $l$, the first $l$ columns of the top half of \eqref{eq:syl1} are filled with zeros. 
Similarly, under condition~3, $n-n^*=l$ for all positive integers $l$, the first~$l$ columns of the bottom half 
of~\eqref{eq:syl1} are filled with zeros. Nevertheless, in all three cases,~\eqref{eq:syl1} does not lose rank since 
it is guaranteed that there is at least one non-zero entry in each column due to the non-monic 
model denominator assumption. Therefore, $S(-B^*(p),A^*(p))$ is non-singular under 
Assumption~\ref{assum5}.

Now, the regressor vector in~\eqref{eq:phif2} can be written as
\begin{equation}	\label{eq:regressor}
	\varphi_f(t_k) = \frac{S(-B^*(p),A^*(p))}{A_j(p)A^*(p)}U_{du}
	- \frac{1}{A_j(p)} P_{reg}.
\end{equation}

Similarly, the instrument vector in \eqref{eq:srivc} can be written as
\begin{equation} \label{eq:ins} 
	\hat{\varphi}_f(t_k) =\frac{S(-B_j(p),A_j(p))}{A_j^2(p)} U_{du},
\end{equation}
where $S(-B_j(p),A_j(p))$ is an $(n+m+1)\times(n+m+1)$ Sylvester matrix defined in the same way as 
\eqref{eq:syl1} with $a_1^*,\dots,a_{n^*}^*,b_0^*,\dots,b_{m^*}^*$ replaced by 
$a_1,\dots,a_n,b_0,\dots,b_m$. From now on, we omit the argument $p$ in the Sylvester 
matrices for simplicity of notation.

It has been shown \cite{Soderstrom1975} that, as $N \rightarrow \infty$, the sums in~\eqref{eq:srivc} 
can be replaced by their expectations, provided $\hat{\varphi}_f(t_k)$, $\varphi_f^\top(t_k)$ and $y_f(t_k)$ 
are jointly stationary stochastic processes. Now, substituting \eqref{eq:regressor} and \eqref{eq:ins} into the 
matrix inverse term in \eqref{eq:srivc}, we obtain
\begin{align} \label{eq:norm1}
	&E\left\{\hat{\varphi}_f(t_k) \varphi_f^\top(t_k)\right\}   \notag \\
	&\;\;\;\;\;\;\;\;\;\;\;=S(-B_j,A_j)\Phi S^\top(-B^*,A^*) -S(-B_j,A_j)\Psi,
\end{align}
where
\begin{equation}	\label{eq:P_mat}
	\Phi = E\left\{\frac{1}{A_j^2(p)} U_{du} \frac{1}{A_j(p)A^*(p)} U^\top_{du}\right\},
\end{equation}
and 
\begin{equation}	\label{eq:P1_mat}
	\Psi = E\left\{\frac{1}{A_j^2(p)}U_{du}\frac{1}{A_j(p)} P^\top_{reg}\right\}.
\end{equation}
According to Lemma A3.1 in \cite{Soderstrom1983}, the Sylvester matrices 
$S(-B_j,A_j)$ and~$S(-B^*,A^*)$ are non-singular provided that $B_j(p)$ and $A_j(p)$ are coprime
and $B^*(p)$ and $A^*(p)$ are coprime. 
For \eqref{eq:norm1} to be non-singular, it is sufficient to show that $\Phi$ is non-singular, 
and $\Psi=0$.


Consider $\Psi$ in \eqref{eq:P1_mat}. An arbitrary entry in the first $n$ columns of $\Psi$ can be written 
in the form of
\begin{align} \label{eq:Psi_freq}
	\Psi_{il} &=E\left\{\frac{p^{n+m+1-i}}{A_j^2(p)}u(t_k) \frac{p^{n+1-l}}{A_j(p)}v(t_k)\right\}	\notag \\
			&= \frac{1}{2\pi} \int_{-\pi}^{\pi}	 \frac{\tilde{B}_i(e^{j\omega})}{\tilde{A}_j^2(e^{j\omega})}
			\frac{\tilde{D}_l(e^{-j\omega})}{\tilde{A}_j(e^{-j\omega})} \phi_{uv}(\omega) d\omega,
\end{align}
where $\tilde{B}_i/\tilde{A}_j^2$ and $\tilde{D}_l/\tilde{A}_j$ are the FOH equivalents of their CT 
transfer functions respectively, $i = 1,\dots,n+m+1$, $l = 1,\dots,n$, and $\phi_{uv}(\omega)$ is the cross-spectrum 
of $u(t_k)$ and $v(t_k)$. Since the input and noise are uncorrelated, $\phi_{uv}(\omega)=0$. Thus, $\Psi=0$.

Now, consider $\Phi$ in \eqref{eq:P_mat}. Similarly, an arbitrary entry of this matrix can be written as
\begin{align} \label{eq:Phi_freq}
	\Phi_{il} &=E \left\{  \frac{p^{n+m+1-i}}{A_j^2(p)}u(t_k)  \frac{p^{n+m+1-l}}{A_j(p)A^*(p)}
			u(t_k)\right\} \notag \\
			&= \frac{1}{2\pi} \int_{-\pi}^\pi \frac{\tilde{B}_i(e^{j\omega})}{\tilde{A}_j^2
			(e^{j\omega})} \frac{\tilde{B}_l(e^{-j\omega})}{\tilde{A}_j(e^{-j\omega})\tilde{A}^*
			(e^{-j\omega})} dF_u(\omega),
\end{align}
where $i,l = 1,\dots,n+m+1$, and $F_u(\omega)$ is the spectral distribution of $u(t_k)$.

We have shown in Lemma~\ref{lem:nonsingularity1} (see Appendix) that 
$\Phi$ is positive definite when evaluated at the true system parameters.
By Lemma~\ref{lem:nonsingularity3} (see Appendix), we have also shown that for a fixed input signal, every entry 
of $\Phi$ is an analytic function of the model parameters. 
Hence, by Lemma A2.3 
of~\cite{Soderstrom1983}, we can conclude that $\Phi$ is generically non-singular. 
Since $S(-B_j,A_j)$ and $S(-B^*,A^*)$ are non-singular, and $\Psi=0$, 
$E\{\hat{\varphi}_f(t_k) \varphi_f^\top(t_k)\}$ is generically non-singular.
\;\;\;\;\;\;\;\;\;\;\;\;\;\;\;\;\;\;\;\;\;\;\;\;\;\;\;\;\;\;\;\;\;\;\;\;\;\;\;\;\;\;\;\;\;\;\;\;\;\;\;\;\;\;\;\;\;\;\;\qed
\end{pf*}
\vspace{-0.7cm}
\begin{pf*} {\textit{Proof of Theorem~\ref{consistency}, Statement 2.}}
Here we will show that, upon convergence, the limiting point of the SRIVC estimator corresponds to
the true parameters. Suppose $\bar{\theta}$ is a limiting point of the iteration in~\eqref{eq:srivc}, and 
the corresponding polynomials of the model are given by
\begin{equation*}
\begin{split}
	\bar{B}(p) &= \bar{b}_0p^{m} + \bar{b}_1p^{m-1} + \cdots + \bar{b}_m	\\
	\bar{A}(p) &= \bar{a}_1p^{n} + \bar{a}_2p^{n-1} + \dots + \bar{a}_np + 1.
\end{split}
\end{equation*}
The polynomials $\bar{B}(p)$ and $\bar{A}(p)$ are coprime since $\bar{\theta}$ satisfies the conditions in 
Statement 1, and one of the conditions is that the Sylvester matrix is non-singular.
Now, at the converging point $\bar{\theta}$, as $N\rightarrow \infty$, the SRIVC 
expression in~\eqref{eq:srivc} implies that
\begin{align}		\label{eq:limiting}
	E\left\{\hat{\varphi}_f(t_k,\bar{\theta}) \varphi_f^\top(t_k,\bar{\theta}) \right\}^{-1} 
	 E\left\{\hat{\varphi}_f(t_k,\bar{\theta}) \varepsilon(t_k,\bar{\theta}) \right\} = 0,
\end{align}
where $\varepsilon(t_k,\bar{\theta})$ is the GEE evaluated at the converging point.
Since the matrix inverse in~\eqref{eq:limiting} is non-singular by Statement~1, the second expectation 
in~\eqref{eq:limiting} must be zero, i.e.
\begin{equation}	\label{eq:r2_res}
	E\Big{\{}\hat{\varphi}(t_k,\bar{\theta}) \varepsilon(t_k,\bar{\theta}) \Big{\}} = 0.
\end{equation}
The GEE in \eqref{eq:gee} can be rearranged as
\begin{align} \label{eq:r2_eqerror}
	\varepsilon(t_k,\bar{\theta}) 
	= \frac{1}{\bar{A}(p)A^*(p)}[\bar{A}(p)B^*(p)&-\bar{B}(p)A^*(p)]u(t_k) \notag \\
	&\;\;\;\;\;\;\;\;\;\;\;\;+ v(t_k). 
\end{align}
Let $\bar{A}(p)B^*(p)-\bar{B}(p)A^*(p) = h_0p^{r} + h_1p^{r-1} + \cdots + h_{r}$, where
$r = \max(n+m^*,n^*+m)= n+m$.
Then, the GEE can be expressed as
\begin{equation}		\label{eq:r2_err}
	\varepsilon(t_k,\bar{\theta}) = \frac{1}{\bar{A}(p)A^*(p)} \left[\begin{array}{ccc}
	u^{(n+m)}(t_k) & \dots & u(t_k)
	\end{array}\right] H + v(t_k),
\end{equation}
where
\begin{equation}	\label{eq:H}
	H = \left[\begin{array}{cccc}
	h_0 & h_1 & \dots & h_{n+m}
	\end{array}\right]^\top.
\end{equation}
Now, substituting \eqref{eq:ins} for $\hat{\varphi}(t_k,\bar{\theta})$ and~\eqref{eq:r2_err} for 
$\varepsilon(t_k,\bar{\theta})$ into~\eqref{eq:r2_res}, we obtain
\begin{equation}	\label{eq:r2_res2} 
	E\Big{\{}\hat{\varphi}(t_k,\bar{\theta}) \varepsilon(t_k,\bar{\theta}) \Big{\}} =
	S(-\bar{B},\bar{A})\bar{\Phi} H + S(-\bar{B},\bar{A}) \tilde{\Psi},
\end{equation}
where $\bar{\Phi}$ is \eqref{eq:P_mat} evaluated at the converging point, and
\begin{equation*}
	\tilde{\Psi} = E\left\{\frac{1}{\bar{A}^2(p)}U_{du} v(t_k)\right\}.
\end{equation*}
By following the same procedure as the proof of Statement 1, we can show that $\bar{\Phi}$ is generically 
non-singular and $\tilde{\Psi}=0$. Thus, for~\eqref{eq:r2_res2} to be zero, $H=0$, which implies
\begin{equation*}
\begin{split}
	\bar{A}(p)B^*(p) &-\bar{B}(p)A^*(p) = 0	\\
	\frac{\bar{B}(p)}{\bar{A}(p)} &= \frac{B^*(p)}{A^*(p)},
\end{split}
\end{equation*}
i.e. $\theta^*$ is the unique limiting point.
\;\;\;\;\;\;\;\;\;\;\;\;\;\;\;\;\;\;\;\;\;\;\;\;\;\;\;\qed
\end{pf*}
\vspace{-0.7cm}
\begin{pf*} {\textit{Proof of Theorem~\ref{consistency}, Statement 3.}}
Let $\bar{\theta}$ be the limiting point, then, as $N\rightarrow \infty$,
\begin{equation*}
	\theta_{j+1} - \bar{\theta} =f_1(\theta_j)f_2(\theta_j),
\end{equation*}
where
\begin{equation*}
	f_1(\theta_j) = E\left\{\hat{\varphi}_f(t_k,\theta_j) \varphi_f^\top(t_k,\theta_j) \right\}^{-1},
\end{equation*}
and
\begin{equation*}
	f_2(\theta_j) = E\left\{ \hat{\varphi}_f(t_k,\theta_j) (y_f(t_k,\theta_j) - 
	\varphi_f^\top(t_k,\theta_j)\bar{\theta}) \right\}.
\end{equation*}
To examine how the SRIVC estimate behaves around the limiting point, we can
linearise $\theta_{j+1}$ around $\bar{\theta}$ using a first order Taylor series, i.e.
\begin{equation*}
\begin{split}
	\theta_{j+1}-\bar{\theta} \approx f_1(\bar{\theta})f_2(\bar{\theta}) &+ \Big{(}
	\left.\frac{\partial f_1(\theta_j)}{\partial \theta_j}\right|_{\theta_j=\bar{\theta}} f_2(\bar{\theta}) \\
	&+f_1(\bar{\theta})\left.\frac{\partial f_2(\theta_j)}{\partial \theta_j}\right|_{\theta_j=\bar{\theta}} \Big{)}
	(\theta_j - \bar{\theta}).
\end{split}
\end{equation*}
At the limiting point, $f_2(\bar{\theta})=0$ as given by~\eqref{eq:r2_res}.
Hence,
\begin{equation}	\label{eq:r3_conv}
	\theta_{j+1}-\bar{\theta} \approx
	f_1(\bar{\theta})\left.\frac{\partial f_2(\theta_j)}{\partial \theta_j}\right|_{\theta_j=\bar{\theta}}
	(\theta_j - \bar{\theta}),
\end{equation}
where 
\begin{align}	\label{eq:df2}
	&\left.\frac{\partial f_2(\theta_j)}{\partial \theta_j}\right|_{\theta_j=\bar{\theta}} 	
	\hspace{-0.5cm}
	=E\left\{ \left.\frac{\partial \hat{\varphi}_f(t_k,\theta_j)}{\partial \theta_j}\right|_{\theta_j=\bar{\theta}}
	\hspace{-0.4cm}
	(y_f(t_k,\bar{\theta})-\varphi_f^\top(t_k,\bar{\theta})\bar{\theta})\right\}	\notag \\
	&+ E\left\{ \hat{\varphi}_f(t_k,\bar{\theta}) \left(
	\left.\frac{\partial y_f(t_k,\theta_j)}{\partial \theta_j}\right|_{\theta_j=\bar{\theta}}  
	\hspace{-0.5cm} -
	\left.\frac{\partial \varphi_f^\top(t_k,\theta_j)}{\partial \theta_j}\right|_{\theta_j=\bar{\theta}}
	\hspace{-0.2cm}\bar{\theta}\right)\right\} \notag \\
	&\;\;\;\;\;\;\;\;\;\;\;\;\;\;\;\;\;=\Psi_1 + \Psi_2.
\end{align}
After some vector differentiations and substituting the expression
\begin{equation*}
\begin{split}
	y_f(t_k,\bar{\theta}) - \varphi_f^\top(t_k,\bar{\theta})\bar{\theta} &= 
	y_f(t_k,\theta^*) - \varphi_f^\top(t_k,\theta^*)\theta^*\\ 
	&=v(t_k)
\end{split}
\end{equation*}
into~\eqref{eq:df2}, we can express $\Psi_1$ as
\begin{equation*}	
	\Psi_1 = E\left\{Mu(t_k)v(t_k)\right\},
\end{equation*}
where
\begin{equation}	\label{eq:dev_M}
M= 
\begin{bmatrix}
\dfrac{p^n}{\bar{A}^3(p)}\left[ 
        -2p^n\bar{B}  \cdots   -2p\bar{B}  \;\;  p^m\bar{A}  \;\cdots\; \bar{A}	
	\right] \\
\vdots \\
\dfrac{p}{\bar{A}^3(p)}\left[
        -2p^n\bar{B}   \cdots   -2p\bar{B}  \;\;  p^m\bar{A}    \;\cdots\;   \bar{A}	
    	\right]   \\
\dfrac{p^m}{\bar{A}^2(p)}\left[
        -p^n   \cdots   -p  \;   0	    \;\cdots\;     0	
    	\right] \\
\vdots \\
\dfrac{1}{\bar{A}^2(p)}\left[
        -p^n   \cdots   -p  \;   0	    \;\cdots\;     0	
    	\right]	\\
\end{bmatrix}^\top.
\end{equation}
Similar to the procedure undertaken in Statement 1, each element of $\Psi_1$ can be expressed as
\begin{equation*}
	 \frac{1}{2\pi} \int_{-\pi}^{\pi}\tilde{G}(e^{j\omega}) \phi_{uv}(\omega) d\omega
\end{equation*}
where $\tilde{G}$ represents the FOH equivalent of the transfer functions in \eqref{eq:dev_M}. 
Since the input and noise are uncorrelated, $\Phi_{uv}=0$. Hence, $\Psi_1=0$.

Now, after some further vector differentiations, $\Psi_2$ in~\eqref{eq:df2} can be written as
\begin{equation*}
\begin{split}
	\Psi_2 &= -S(-\bar{B},\bar{A})E\left\{\frac{1}{\bar{A}^2(p)}U_{du}\frac{1}{\bar{A}(p)}P_{reg}^\top \right\}\\
	&=-\bar{\Psi} \\
	&=0,
\end{split}
\end{equation*}
where $\bar{\Psi}$ is \eqref{eq:P1_mat} evaluated at the converging point. Hence, \eqref{eq:df2} is equal to zero. 
Therefore, according to \eqref{eq:r3_conv}, $\theta_{j+1}$ asymptotically converges to $\theta^*$ for $j \geq 1$, 
and this completes the final part of the proof.
\;\;\;\;\;\;\;\;\;\;\;\;\;\;\;\;\;\;\;\;\;\;\;\;\;\;\;\;\qed
\end{pf*}




\begin{cor} \label{cor_zoh}
When the FOH used in Theorem~\ref{consistency} is replaced with a ZOH, and the intersample behaviour of 
the true system input $\mathring{u}(t)$ satisfies Assumption~\ref{assum6}, statements~1, 2 and 3 in 
Theorem~\ref{consistency} still hold. 
\end{cor}
\begin{pf*} {\textit{Proof of Corollary~\ref{cor_zoh}.}}
The proof follows the same procedure as that shown in Theorem~\ref{consistency}.
Note that, in this case, when the system and model transfer functions are strictly proper, 
the numerator degree of the DT transfer function is at most $n-1$. Thus, 
the persistent excitation order in Assumption~\ref{assum3} can be relaxed to $2n$ 
according the reasoning provided in Remark~\ref{rem_pe} (see Appendix).
\qed
\end{pf*}


Theorem~\ref{consistency} and Corollary~\ref{cor_zoh} have established consistency when the intersample 
behaviours of the input signals in both the regressor and instrument vectors as well as the output signal are 
assumed to be the same as that of the true system input.
Next, Corollary~\ref{corr_input_intersample} examines the effect on the consistency of the SRIVC estimates 
when an incorrect intersample behaviour is assumed for 1)~the input in the instrument vector, and 2)~the 
input in the regressor vector (the model input).
Again, the true system input $\mathring{u}(t)$ is assumed to have an FOH for discretisation purposes, and 
an \emph{incorrect} intersample behaviour means that the signal has an intersample behaviour that is 
different from the true system input.

\begin{cor} \label{corr_input_intersample}
The SRIVC estimator;
\begin{enumerate}
	\item remains generically consistent if an incorrect assumption on the intersample 
	behaviour is used 
	for generating the filtered signals in the instrument vector~$\hat{\varphi}_f(t_k)$ (this includes the generation 
	of the noise-free model output $x(t_k)$, the filtered noise-free model output $x_f(t_k)$, and the filtered input 
	signal $u_f(t_k)$ in~\eqref{eq:srivc_ins}); 
	and 
	\item is generically not consistent if an incorrect assumption on the intersample behaviour is 
	used for filtering the input signal in the regressor vector~$\varphi_f(t_k)$.
\end{enumerate}
\end{cor}
\begin{pf*} {\textit{Proof of Corollary~\ref{corr_input_intersample}, Statement 1.}}
The input used to form the instrument vector is assumed to have a ZOH. Statement~1 of 
Theorem~\ref{consistency} still holds since the only change is that the FOH discretisations of the first 
transfer functions in \eqref{eq:Psi_freq} and \eqref{eq:Phi_freq} are replaced by their ZOH equivalents, 
and this does not affect the way analyticity of $\Psi$ is shown.
Statement~2 in Theorem~\ref{consistency} remains unchanged since the incorrect intersample behaviour 
assumption of the input in the instrument does not affect the formulation of the equation error 
in~\eqref{eq:r2_eqerror}. 
For the same reason, Statement~3 of 
Theorem~\ref{consistency} also remains unchanged. Therefore, when an incorrect intersample behaviour 
for the input signal in the instrument vector $\hat{\varphi}_f(t_k)$ is assumed, the SRIVC estimator remains 
generically consistent.
\;\;\;\;\;\;\;\;\;\;\;\;\;\;\;\;\;\;\;\;\;\;\;\;\;\;\;\;\;\;\;\;\;\;\;\;\;\;\;\;\;\;\;\;\;\;\;\;\;\;\;\;\;\;\qed
\end{pf*}
\vspace{-0.7cm}
\begin{pf*} {\textit{Proof of Corollary~\ref{corr_input_intersample}, Statement 2.}}
Let the input in the regressor vector, indicated by~$\tilde{u}(t_k)$, have a different intersample behaviour 
from that of the true system input. Statement~1 of Theorem~\ref{consistency} remains unchanged, however, 
we will show that Statement 2 of Theorem~\ref{consistency} has been affected. 
Consider the GEE at the limiting solution~$\bar{\theta}$
\begin{align}	\label{eq:cor2_err}
	\varepsilon(t_k,\bar{\theta}) &= \mathring{y}(t_k) - \frac{\bar{B}(p)}{\bar{A}(p)} \tilde{u}(t_k)	\notag \\
	&= \frac{B^*(p)}{A^*(p)}u(t_k) + v(t_k) - \frac{\bar{B}(p)}{\bar{A}(p)} \tilde{u}(t_k)	\notag \\
	&= \frac{B^*(p)}{A^*(p)}u(t_k) + v(t_k) - \left(\frac{\bar{B}(p)}{\bar{A}(p)}u(t_k)+\varepsilon_u(t_k)\right) \notag \\
	&= \frac{1}{\bar{A}(p)A^*(p)}[\bar{A}(p)B^*(p)-\bar{B}(p)A^*(p)]u(t_k) \notag \\
	& \;\;\;\;\;\;\;\;\;\;\;\;\;\;\;\;\;\;\;\;\;\;\;\;\;\;\;\;\;\;\;\;\;\;\;\;\;+ v(t_k) - \varepsilon_u(t_k).
\end{align}
We have introduced an input-dependent term $\varepsilon_u(t_k)$ into the modelled output 
in~\eqref{eq:cor2_err} to account for the interpolation error.
At the limiting point, \eqref{eq:r2_res} holds.
Substituting \eqref{eq:ins} for $\hat{\varphi}_f(t_k,\bar{\theta})$ and~\eqref{eq:cor2_err} for 
$\varepsilon(t_k,\bar{\theta})$ into \eqref{eq:r2_res}. By using the same definition of $H$ from~\eqref{eq:H}, 
we obtain
\begin{align}	\label{eq:cor2_zero2}
	&E\left\{\hat{\varphi}(t_k,\bar{\theta}) \varepsilon(t_k,\bar{\theta}) \right\}  \notag \\
	&\;\;\;\;\;\;\;\;\;=S(-\bar{B},\bar{A}) \bar{\Phi}H - S(-\bar{B},\bar{A})
	E\left\{\frac{1}{\bar{A}^2(p)}U_{du}\varepsilon_u(t_k)\right\} \notag\\
	&\;\;\;\;\;\;\;\;\;=S(-\bar{B},\bar{A}) \bar{\Phi}H - S(-\bar{B},\bar{A})\tilde{\Psi}_u \notag \\
	&\;\;\;\;\;\;\;\;\;=0.
\end{align}
Since the error $\varepsilon_u(t_k)$ is input dependent, $\tilde{\Psi}_u$
does not go to zero in general. The matrix $\bar{\Phi}$ is generically non-singular by Statement~1 
of Theorem~\ref{consistency}. Therefore, we can obtain the coefficients of $H(p)$ by solving
\begin{equation}	\label{eq:H_par}
	H = \left[S(-\bar{B},\bar{A})\bar{\Phi}\right]^{-1}S(-\bar{B},\bar{A})\tilde{\Psi}_u.
\end{equation} 
Now,
\begin{align}	\label{eq:new_theta_bar}
	\bar{A}(p)B^*(p) &-\bar{B}(p)A^*(p) = H(p)	\notag \\
	\frac{\bar{B}(p)}{\bar{A}(p)} &= \frac{B^*(p)}{A^*(p)} + \frac{H(p)}{\bar{A}(p)A^*(p)},
\end{align}
where the parameters of the polynomial $H(p)$ are given by \eqref{eq:H_par}. The expression 
in~\eqref{eq:new_theta_bar} shows that the true 
parameters are no longer the limiting solution of the SRIVC estimator, i.e. $\bar{\theta} \neq \theta^*$.

It is implied by~\eqref{eq:r2_res} that the input is uncorrelated with the GEE evaluated at the converging point.
Therefore, $\theta_{j+1}$ asymptotically converges to the 
new limiting point~$\bar{\theta}$ given in \eqref{eq:new_theta_bar} for $j \geq 1$. Together with 
$\bar{\theta} \neq \theta^*$, we can conclude that $\theta_{j+1}$ does not converge to the true parameters 
$\theta^*$ if an incorrect intersample behaviour is assumed for the input signal in the 
regressor vector $\varphi_f(t_k)$. Hence, the SRIVC estimator is generically not consistent when the true system 
input cannot be interpolated exactly.
\;\;\;\;\;\;\;\;\;\;\;\;\;\;\;\;\;\;\;\;\;\;\;\;\;\;\;\;\;\;\;\;\;\;\;\;\;\;\;\;\;\;\;\;\;\;\;\qed
\end{pf*}

For discretisation purposes, the true system input is assumed to have an FOH for the analysis above. 
We note that Corollary~\ref{corr_input_intersample} holds for any input that cannot be interpolated exactly, 
as stated in the following remark.



\begin{rem} \label{remark_ct_input}
When the input to the real system $\mathring{u}(t)$ is a continuous function of time that cannot be 
interpolated exactly, the SRIVC estimator is not consistent. This follows from 
Corollary~\ref{corr_input_intersample}.
\end{rem}

We note that in situations where the input cannot be interpolated 
exactly, $\varepsilon_u(t_k)$ will be non-zero, and
the bias on the estimates is captured by $\frac{H(p)}{\bar{A}(p)A^*(p)}$ in \eqref{eq:new_theta_bar}. 
Since the polynomial $H(p)$ is proportional to the interpolation 
error $\varepsilon_u(t_k)$, which will decrease if the signals are sampled faster, 
this implies that the bias on the estimates will generally decrease with the sampling period.




Next, we examine the effect of the intersample behaviour of the sampled output on the consistency 
of the SRIVC estimator in the following remark.

\begin{rem} \label{remark_intersample_y}
Consider the GEE $\varepsilon(t_k)$. At each SRIVC iteration $j$, $\varepsilon(t_k)$ can be 
expressed as
\begin{align}	\label{eq:remark_GEE}
	&\varepsilon(t_k,\theta_j)  \notag\\
	&\;\;\;= A_j(p)\left(\frac{1}{A_{j-1}(p)}\mathring{y}(t_k)\right)- B_j(p)\left(\frac{1}{A_{j-1}(p)}u(t_k)\right).
\end{align}

\vspace{-0.8cm}
Upon convergence, the expression in \eqref{eq:remark_GEE} becomes
\begin{equation*}
	\varepsilon(t_k,\bar{\theta}) = \bar{A}(p)\left(\frac{1}{\bar{A}(p)}\mathring{y}(t_k)\right) 
	- \bar{B}(p)\left(\frac{1}{\bar{A}(p)}u(t_k)\right).
\end{equation*}
Hence, the intersample behaviour of the measured output $\mathring{y}(t_k)$ does not affect the GEE as 
the discretisation of $1/\bar{A}(p)$ cancels with that of $\bar{A}(p)$ at the converging point, and thus it 
does not influence the consistency of the SRIVC estimator.

\end{rem}

Note it has been empirically observed that even though the intersample behaviour of $\mathring{y}(t_k)$ does 
not affect the consistency of the SRIVC estimator, a better interpolation of this signal, e.g. using a FOH as 
opposed to a ZOH, can speed up the rate of convergence of the SRIVC iteration.




\section{Simulation Results}	\label{sec:sim}

\begin{figure} [h]
\begin{center}
\includegraphics[width = 8.5cm]{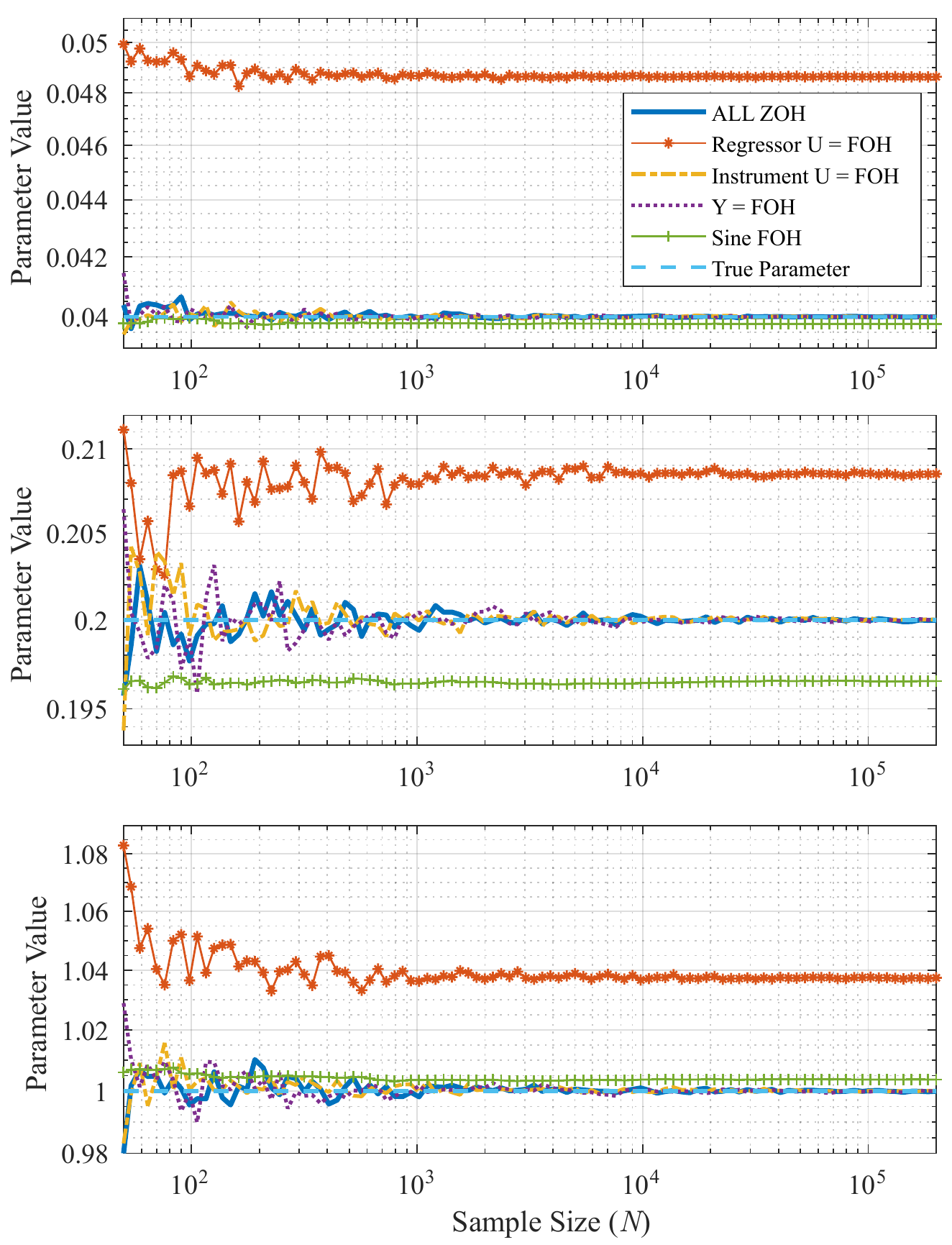}
\caption{Mean of the estimated parameters.}
\label{fig:mean}
\end{center}
\end{figure}

\begin{figure} [h]
\begin{center}
\includegraphics[width = 8.5cm]{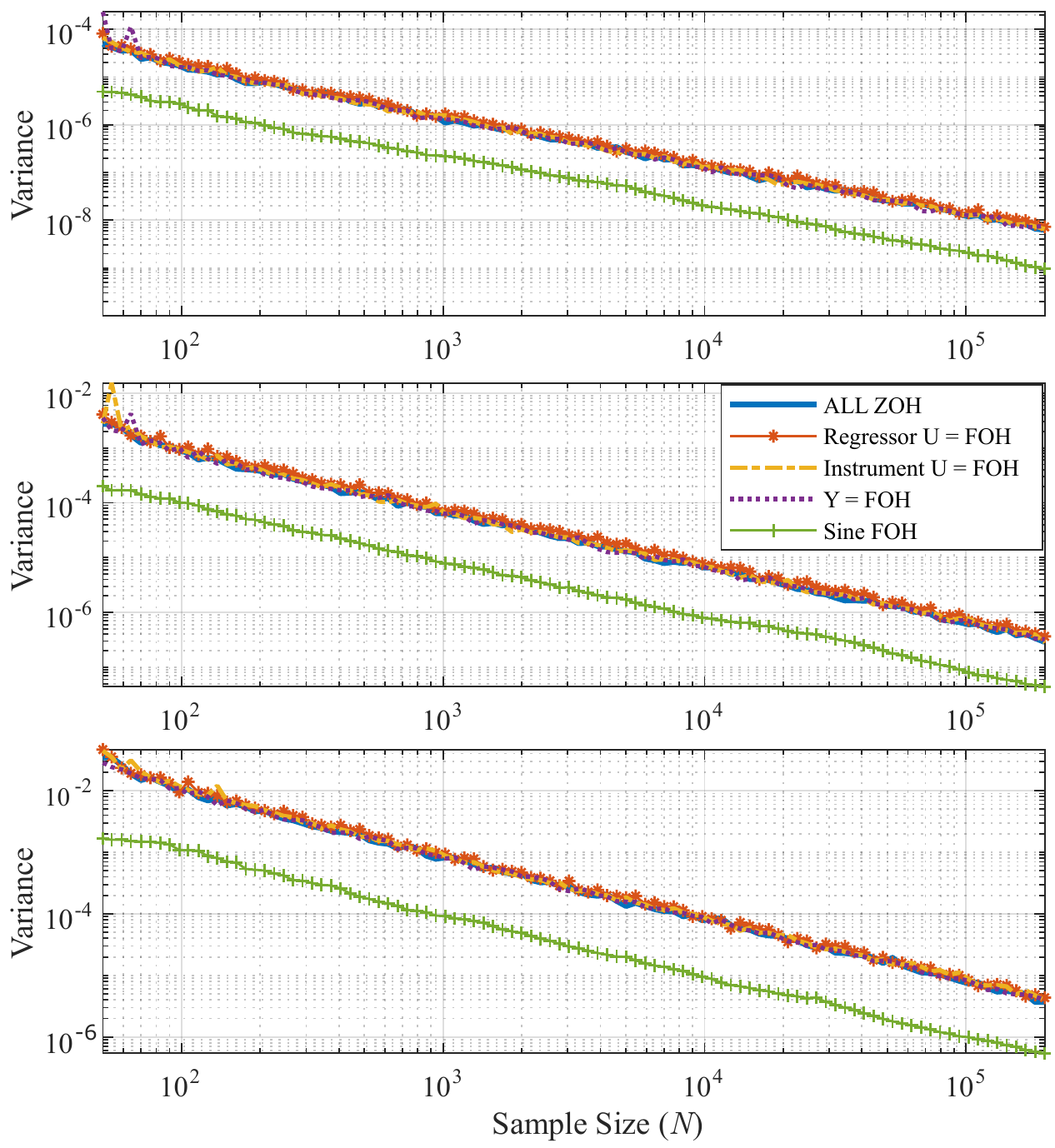}
\caption{Variance of the estimated parameters.}
\label{fig:variance}
\end{center}
\end{figure}

Monte Carlo simulations are performed for a second order system to support the theoretical 
analyses developed in the previous section. The second order system is chosen to be
\begin{equation*}
	G^*(p) = \frac{1}{0.04p^2 + 0.2p + 1},
\end{equation*}
and the true parameters are given by
\begin{equation*}
	\theta^* = \left[ \begin{array}{ccc}
	0.04 & 0.2 & 1
	\end{array} \right]^\top.
\end{equation*}
The measured signals are sampled at $T = 0.1$ s, and the input is chosen to be a random binary signal uniformly 
exciting the system from $0$ Hz up to the Nyquist frequency. The input applied to the true system 
$\mathring{u}(t)$ has a 
zero-order hold intersample behaviour. The additive noise on the output is an i.i.d. Gaussian sequence with 
a variance of $0.1$. The consistency of the SRIVC estimator is investigated by examining the mean and 
variance of the estimates in a Monte Carlo simulation study as the sample size~$N$ increases. Here, $N$ 
is adjusted from $50$ to $200000$ in a logarithmic scale, where a total of $100$ different sample sizes are 
used. Three hundred Monte Carlo simulations are performed for each value of $N$ with the mean and 
variance of the three parameter estimates calculated. The maximum number of iterations of the SRIVC algorithm 
is set to $200$, and the relative error bound $\epsilon$ in \eqref{eq:srivc_stop} is set to $10^{-7}$.
The mean and variance of the estimated parameters 
with respect to an increasing sample size are examined under four different cases by changing the intersample 
behaviour of the measured signals when discretising different filters in the SRIVC algorithm. These cases 
include 
\begin{itemize}
	\item matching the intersample behaviour of all the signals in the algorithm with that of $\mathring{u}(t)$, 
	which is a ZOH, 
	\item setting only the intersample behaviour of $u(t_k)$ in the regressor vector to FOH,
	\item setting only the intersample behaviour of $u(t_k)$ in the instrument vector to FOH, and
	\item setting only the intersample behaviour of $\mathring{y}(t_k)$ to FOH.
\end{itemize}
These cases correspond to the first four instances in Fig.~\ref{fig:mean} and Fig.~\ref{fig:variance}.

In another simulation, a multisine input, given by
\begin{equation*}
	\mathring{u}(t) = \sin(0.5t) + \sin(2t) + \sin(5t) + \sin(7t),
\end{equation*}
is used to excite the true system $G^*(p)$. The noiseless output is computed analytically 
by assuming that it corresponds to the system output at stationary state, i.e.
\begin{equation*}
	x(t) = \sum_{i=1}^{4} |G^*(j\omega_i)|\sin(\omega_i t + \angle G^*(j\omega_i)),
\end{equation*}
where $\{\omega_1,\omega_2,\omega_3,\omega_4\} = \{0.5,2,5,7\}$.
The CT input and output are also sampled at $T=0.1$ s, and the additive noise on the 
measured output is an i.i.d. Gaussian sequence with a variance of $0.1$. The same Monte Carlo studies as 
described previously for the random binary input are performed for the multisine input. The model input 
is interpolated using a FOH to approximate $\mathring{u}(t)$ as close as possible. The mean and variance 
of the estimates for each sample size are calculated to examine the consistency of the SRIVC 
estimator in situations when the input cannot be interpolated exactly. This corresponds to the 
fifth instance in Fig.~\ref{fig:mean} and Fig.~\ref{fig:variance}. 

The mean and variance of the estimated parameters for the five instances described previously are 
shown in Fig.~\ref{fig:mean} and Fig.~\ref{fig:variance} respectively. The three subplots in both figures 
correspond to the parameters in the order of $0.04, 0.2, 1$.
The true parameters are plotted with a dotted line in the three subplots of Fig.~\ref{fig:mean}. 
We can see in Fig.~\ref{fig:variance} that the variance of the SRIVC estimates decreases with an increasing 
sample size in all cases. More oscillations in the mean values are observed for small sample sizes, but the 
estimates eventually converge to the true parameters after approximately $10000$ samples in situations 
where we have matched the intersample behaviour of $u(t_k)$ in the regressor vector with that of the input 
applied to the true system, i.e. instances 1, 3 and 4. Together with the decreasing variance, this provides 
empirical evidence to the consistency result in Theorem~\ref{consistency}. In addition, changing the intersample 
behaviour of the input in the instrument vector or the output does not seem to affect the consistency of the 
SRIVC estimates, which aligns with statement 1 of Corollary~\ref{corr_input_intersample} and 
Remark~\ref{remark_intersample_y} respectively.
We can also see that when the model input does not match the true system input, the estimates 
do not converge to the true parameters with an increasing sample size. The 
SRIVC estimator is not consistent in this case, which has been shown theoretically in statement 2 of 
Corollary~\ref{corr_input_intersample}.
Furthermore, when a CT signal, which cannot be interpolated exactly, is used as the true 
system input, the SRIVC estimator can also be seen to be inconsistent.
We do note that the bias on the SRIVC estimates can be reduced if more sophisticated interpolation 
methods other than a ZOH or FOH are used to reconstruct the input signal, and the bias will decrease 
with a decreasing sampling period as shown in statement 2 of Corollary~\ref{corr_input_intersample}.

\section{Conclusion}		\label{sec:conclusion}

In this paper, we have analysed the consistency of the SRIVC estimator by taking into 
account the intersample behaviour of the input signal, and conducted simulation experiments to provide 
empirical observations to the theoretical results. The main result of the paper is that the SRIVC estimator 
is generically consistent when the 
intersample behaviour of the input signal applied to the continuous-time system is known exactly and 
used adequately in the implementation of the algorithm. 
It has been shown that when the intersample behaviour of the input signal in the regressor, i.e. the model 
input, does not match that of the true system input, the unique converging point of the estimator no 
longer corresponds to the true parameters, and thus the SRIVC estimator is generically not consistent.
On the other hand, the intersample behaviours of the input signal in the instrument vector or the output 
signal in the regressor vector do not affect the consistency of the estimator. 

\section{Appendix}	\label{sec:app}

\begin{lem} \label{lem:nonsingularity2}
The mapping between the CT parameters $(a_1,\dots,a_n)$ and the DT 
parameters $(\alpha_1,\dots,\alpha_n)$ is analytic in $\{(a_1,\dots,a_n)\in \mathbb{C}^n: a_1 \neq 0\}$.
\end{lem}

\begin{pf*} {\textit{Proof of Lemma~\ref{lem:nonsingularity2}.}}
Since $a_1\neq 0$, we can write
\begin{align}
A'(p)&:=p^n + \frac{a_2}{a_1}p^{n-1}+\dots+\frac{a_n}{a_1}p+\frac{1}{a_1} \notag \\
&=p^n + a_1' p^{n-1}+\dots+a_{n-1}' p+a_n' \notag 
\end{align}
where we note that the mapping $(a_1,a_2,\dots,a_n)\to (a_1',a_2',\dots,a_n')$ is analytic for $a_1 \neq 0$. 
Let the state matrix of the CT system $1/A'(p)$ be $\textnormal{A}_c$.
The state matrix of the DT equivalent is then $\textnormal{A}_d= \exp(\textnormal{A}_c T)$, 
where $T$ is the sampling period. The exponential is also analytic in the variables $(a_1',a_2',\dots,a_n')$. 
Finally, we can express $\tilde{A}(z) = \det(zI-\textnormal{A}_d)$. It is known that the coefficients of this 
characteristic polynomial are polynomial expressions in the entries of the matrix $\textnormal{A}_d$ 
(see e.g. \cite{brooks2006}). This implies that $\{\alpha_i\}_{i=1}^n$ are analytic functions of 
the entries of $\textnormal{A}_d$. The lemma then follows from the composition of analytic functions.
\;\;\;\;\;\;\;\;\;\;\;\;\;\;\;\;\;\;\;\;\;\;\;\;\;\;\;\;\;\;\;\;\;\;\;\;\;\;\;\;\;\;\;\;\;\;\;\;\;\;\qed
\end{pf*}


\begin{lem}  \label{lem:nonsingularity1}
Consider a FOH sampling, and Assumptions~\ref{assum1}-\ref{assum7} hold. Then, the matrix
$\Phi$ in~\eqref{eq:P_mat} evaluated at the true system parameters, i.e.
\begin{equation}
	\Phi^* := E\left\{ \frac{1}{{A^*}^2(p)} 
	U_{du} \frac{1}{{A^*}^2(p)}
	U_{du}^\top  \right\}>0,
\end{equation}
is positive definite.
\end{lem}

\begin{pf*}{\textit{Proof of Lemma~\ref{lem:nonsingularity1}.}}
Let $\textbf{z}\in \mathbb{R}^{n+m+1}$. We can write
\begin{equation} \label{eq2}
	\textbf{z}^\top \Phi^* \textbf{z} = E\left\{ \left(\frac{B_\textbf{z}(p)}{{A^*}^2(p)}u(t_k)\right)^2 \right\} \geq 0
\end{equation}
where $B_\textbf{z}(p)$ is an arbitrary polynomial of degree $n+m$. In the frequency domain, 
\eqref{eq2} can be written as
\begin{equation}
	\textbf{z}^\top \Phi^* \textbf{z} = \frac{1}{2\pi} \int_{-\pi}^\pi \left| 
	\frac{\tilde{B}_\textbf{z}(e^{j\omega})}{{\tilde{A}^{*^2}}(e^{j\omega})}\right|^2 dF_u(\omega)
\end{equation}
where $\tilde{B}_\textbf{z}$ and $\tilde{A}^{*^2}$ are the FOH equivalent polynomials of 
$B_\textbf{z}(p)/{A^*}^2(p)$, and $F_u(\omega)$ is the spectral distribution of $\{u(t_k)\}$. 
Note that $\tilde{B}_\textbf{z}$ is in general a $2n$ degree polynomial. 

We can also write \eqref{eq2} as
\begin{equation}
	\textbf{z}^\top \Phi^* \textbf{z} = \frac{1}{2\pi} \int_{-\pi}^\pi 
	|\tilde{B}_\textbf{z}(e^{j\omega})|^2 dF_{\tilde{u}}(\omega)
\end{equation}
where the support of the spectral distribution function $\phi_{\tilde{u}}$ consists of at least $2n+1$ points since 
filtering $u(t_k)$ by $1/\tilde{A}^{*^2}$ gives a signal which is also persistently exciting of order at least $2n+1$. 
By the definition of persistence of excitation, $\textbf{z}^\top \Phi^* \textbf{z}=0$ implies 
$\tilde{B}_\textbf{z}(e^{j\omega})\equiv 0$~\cite[Theorem 1]{ljung1971} and hence $\textbf{z}$ is a zero vector. 
This means that $B_\textbf{z}(p)/A^{*^2}(p)$ gives a sampled model equal to zero at all sampling instants. 

Now, assume that there exists a $\textbf{z}_*$ 
such that $\frac{B_{\textbf{z}_*}(p)}{{A^*}^2(p)}u(t_k)=0$. Thus, by linearity, for all $\textbf{z}$ we have
\begin{equation}
	\frac{B_{\textbf{z}}(p)}{{A^*}^2(p)}u(t_k)=\frac{B_{\textbf{z}}(p)+B_{\textbf{z}_*}(p)}{{A^*}^2(p)}u(t_k).
\end{equation}
This means that, if $B_{\textbf{z}_*}(p)$ were non-zero, the CT model is not uniquely determined 
by the DT model. However, this is not true under the  sampling condition of the statement 
(see \cite{kollar1996}). Therefore, it is not possible for a polynomial $B_{\textbf{z}}(p)$ different from zero to give 
a sampled model equal to zero. This means that~\eqref{eq2} is strictly positive for any non-zero vector 
$\textbf{z}$. Hence, $\Phi^*$ is positive definite. 
\qed
\end{pf*}


\begin{rem} \label{rem_pe}
Note that even though there are $n+m+1$ parameters in the CT transfer function to be identified, 
the input is required to have a persistent excitation of order $2n+1$ instead of $n+m+1$ for a first order hold 
discretisation of the input. The reason behind this is that the numerator polynomial $B_\textbf{z}(p)$ gets mapped 
to a subset of a larger space, namely a subset of the space of $2n$ polynomials. Hence, singularity of~$\Phi^*$ 
can be obtained by an unfortunate choice of the frequency lines of $\{u(t_k)\}$. If  
the input is persistently exciting of an order less than $2n+1$, the frequency lines of the input could match the 
zeros of $\tilde{B}_\textbf{z}(e^{j\omega})$, leading to $\textbf{z}^\top \Phi^* \textbf{z}=0$ when $\textbf{z}\neq 0$.
\end{rem}




\begin{lem} \label{lem:nonsingularity3}
Each element of the matrix $\Phi$ in~\eqref{eq:P_mat} is an analytic function of 
$a_1,\dots,a_n$ for $(a_1,\dots,a_n)\in \Omega$, where $\Omega$ denotes the subset of 
$\mathbb{C}^{n}$ consisting of parameter vectors $(a_1,\dots,a_n)$ 
such that $A_j(p)$ has all zeros strictly in the left half-plane.
\end{lem}

\begin{pf*}{\textit{Proof of Lemma~\ref{lem:nonsingularity3}.}}
Define the FOH equivalent of the model denominator as
\begin{equation*}
	\tilde{A}_j(q) := q^{-n} + \alpha_1 q^{-n+1} + ... + \alpha_n,
\end{equation*}
and denote $\Omega_d$ as the subset of $\mathbb{C}^{n}$ 
consisting of parameter vectors $(\alpha_1,\dots,\alpha_n)$ such that $\tilde{A}_j(q)$ has all zeros strictly inside the 
unit circle. By Lemma~\ref{lem:nonsingularity2}, 
there is an analytic mapping between $(a_1,\dots,a_n)$ and the DT parameter vector 
$(\alpha_1,\dots,\alpha_n)$. 
Now, fixing $\alpha_2,\dots,\alpha_n$ allows us to define a region
$\Omega_{d1}\subset \mathbb{C}$ where $\bar{\alpha}_1\in \Omega_{d1}$ implies 
$(\bar{\alpha}_1,\alpha_2,\dots,\alpha_n) \in \Omega_d$. Note that the integrand in \eqref{eq:Phi_freq} is an analytic
 function of $\alpha_1$ in $\Omega_{d1}$, and from now on, we denote this integrand as $f(\bar{\alpha}_1,\omega)$. 

Let $C$ be a closed contour in $\Omega_{d1}$ such that $\bar{\alpha}_1$ is interior to $C$. Then,
\begin{equation}
	f(\bar{\alpha}_1,\omega) = \frac{1}{2\pi i} \int_{C} \frac{f(\alpha_1,\omega)}{\alpha_1-\bar{\alpha}_1} 
	d\alpha_1.
\end{equation}
As a result,
\begin{equation}
	\Phi_{il}(\bar{\alpha}_1) = \frac{1}{2\pi} \int_{-\pi}^\pi \frac{1}{2\pi i} \int_{C} 
	\frac{f(\alpha_1,\omega)}{\alpha_1-\bar{\alpha}_1} d\alpha_1 dF_u(\omega).
\end{equation}
Since the function being integrated is bounded on $[-\pi,\pi]\times C$, the order of integration can be changed by 
Fubini's Theorem~\cite[p. 961]{Stewart2008}, which yields
\begin{align}
	\Phi_{il}(\bar{\alpha}_1) &= \frac{1}{2\pi i} \int_C \frac{1}{\alpha_1-\bar{\alpha}_1} \frac{1}{2\pi} 
	\int_{-\pi}^\pi f(\alpha_1,\omega)dF_u(\omega) d\alpha_1 \notag \\
	&= \frac{1}{2\pi i} \int_C \frac{\Phi_{il}(\alpha_1)}{\alpha_1-\bar{\alpha}_1} d\alpha_1
\end{align}
from which we conclude that $\Phi_{il}(\alpha_1)$ is analytic in a neighbourhood around 
$\alpha_1 = \bar{\alpha}_1$.

Repeating this process for every $\alpha_i$, $i=2,\dots,n$, we obtain that $\Phi_{il}$ is an analytic function of the 
variables $\alpha_1,\alpha_2,\dots,\alpha_n$ separately. Since $\Phi_{il}$ is a continuous function of 
$(\alpha_1,\dots,\alpha_n)$ in $\Omega_d$, $\Phi_{il}$ is an analytic function of the joint variables 
$(\alpha_1,\dots,\alpha_n)$ by Osgood's Lemma \cite[p. 139]{bochner1948}. Hence, each element of $\Phi$ is an 
analytic function of $a_1, \dots, a_n$ for $(a_1, \dots, a_n) \in \Omega$. 
\;\;\;\;\;\;\;\;\;\;\;\;\;\;\;\;\;\;\;\;\;\;\;\;\;\;\;\;\qed
\end{pf*}

\balance
\bibliographystyle{plain}        
\bibliography{library}

\end{document}